# Long-lived transmons with different electrode layouts

Kungang Li,[1,2,3] S. K. Dutta,[1,2,3] Zachary Steffen,[1,2,3,4] Dylan Poppert,[1,2,3] Shahriar Keshvari,[1,2,3] Jeffery Bowser,[1,2,3] B. S. Palmer,[1,2,3,4] C. J. Lobb[1,2,3] and F. C. Wellstood[1,2,3]

1. Department of Physics, University of Maryland, College Park, MD 20742, USA
2. Joint Quantum Institute, University of Maryland, College Park, MD 20742, USA
3. Quantum Materials Center, University of Maryland, College Park, MD 20742, USA
4. Laboratory for Physical Sciences, College Park, MD 20740, USA

## Abstract

To test the contribution of non-equilibrium quasiparticles to qubit relaxation, we have repeatedly measured the relaxation time $T_1$ in Al/AlO$_x$/Al transmons with electrodes that have different superconducting gaps. In one device, the first layer electrode was formed by thermal evaporation of nominally pure Al, while the counter-electrode was formed by deposition of oxygen-doped Al, which gave a larger gap value. The relaxation time was long, but showed large fluctuations, with $T_1$ varying between about 100 and 300 μs at 20 mK. In other transmons, we formed the first layer electrode by deposition of oxygen-doped Al, while the counter-electrode was formed by deposition of nominally pure Al. These devices showed a similar range of large and fluctuating $T_1$ values, with maximum $T_1$ values over 200 μs. The relaxation time of the devices did not depend strongly on temperature below about 150 mK, but dropped rapidly above this due to thermally-generated quasiparticles.

# Introduction

Over the past two decades, better materials and better qubit designs have led to great improvements in the performance of superconducting qubits [1,2]. In particular, the relaxation time $T_1$ has increased by orders of magnitude. In addition, relatively large fluctuations in $T_1$ have also been reported [3] and this may provide information about the dominant relaxation mechanism. It is now well understood that significant relaxation in transmons can be caused by a variety of mechanisms, including dielectric loss due to two-level systems in the substrate or oxide layers [4-6], tunneling of non-equilibrium quasiparticles through the transmon junction [7-9], and the Purcell effect or coupling to lossy microwave modes.

Typically in transmons fabricated from aluminum, the relaxation time decreases rapidly above about 150 mK due to thermally-generated quasiparticles. At lower temperatures, non-equilibrium quasiparticles (generated by the absorption of high energy phonons, photons, cosmic rays or some other energetic source) can be present at sufficiently high densities to contribute significantly to loss, at least in some devices. If quasiparticle tunneling through the transmon's junction is a significant contribution to the loss, then increasing the difference between the superconducting energy gap of the electrode and counter-electrode may increase the relaxation time. In this situation, the non-equilibrium quasiparticles will tend to thermalize and accumulate in the electrode with the lower superconducting energy gap. If the gap difference is larger than the energy difference $hf_{ge}$ between the transmon's ground state and excited states, then quasiparticle tunneling from the low-gap side to the high-gap side will be suppressed [10-13], yielding a longer relaxation time.

In this paper, we discuss our initial results on measurements of the relaxation time of

Al/AlO$_x$/Al 3D transmons that have been fabricated with electrodes that have different superconducting gaps. We first briefly describe our fabrication process and experimental setup. We then present our measurements of $T_1$ as a function of time and temperature and conclude with a discussion of our main results.

**Fabrication of transmons and experimental arrangement**

To produce thin film electrodes with different superconducting gaps, we deposited aluminum in low-pressure O$_2$ [14-19]. The oxygen causes the Al to grow with very small grains and increases the superconducting gap. We first performed test depositions of thermally evaporated Al thin films on sapphire substrates in different pressures of O$_2$. After deposition, the resistance of the films was measured versus temperature to determine the transition temperature $T_c$. As expected, higher oxygen doping pressure and thinner films produced higher $T_c$ values. From $T_c$, we determined the zero-temperature superconducting gap $\Delta$ using the BCS result [20]

$$\Delta = 1.76 k_B T_c . \qquad (1)$$

For O$_2$ pressures in the µTorr range, we were able to obtain gaps in the 200-250 µeV range, which is about 15% to 40% larger than the gap of bulk pure aluminum.

The first two devices we made, $Q_{L1}$ and $Q_{R1}$, were fabricated on the same sapphire chip using electron-beam lithography and double angle evaporation. Following lithography, we used thermal evaporation at one angle to deposit a 28 nm thick layer of nominally pure Al. This formed the electrode layer for both transmons. This layer was exposed to 2 Torr of O$_2$ for 7.5 minutes to grow the AlO$_x$ tunnel barrier. The substrate was then tilted to a second angle, the right half of the chip was covered with a shutter, and a 77 nm thick layer of Al was thermally evaporated in 2.5 µTorr of O$_2$. This layer formed the counter-electrode of device $Q_{L1}$. The shutter was then adjust-

| | Gap-engineered transmon $Q_{L1}$ | Standard transmon $Q_{R1}$ | Gap-engineered transmon $Q_{L2}$ | Gap-engineered transmon $Q_{R2}$ |
|---|---|---|---|---|
| $f_{ge}$ (GHz) | 2.930 | 3.774 | 2.610 | 2.879 |
| $E_J/h$ (GHz) | 5.55 | 9.95 | 4.29 | 5.59 |
| $E_C/h$ (MHz) | 224 | 198 | 236 | 214 |
| Base electrode gap (μeV) | 200.0 | 200.0 | ~250 | ~250 |
| Counter-electrode gap (μeV) | 227.7 | 191.1 | ~200 | ~200 |
| Gap difference (μeV) | 27.7 | 8.9 | ~50 | ~50 |
| $hf_{ge}$ (μeV) | 12.1 | 15.6 | 10.8 | 11.9 |
| Is gap difference > $hf_{ge}$? | Yes | No | Yes | Yes |

**Table 1** Table of transmon parameters. $f_{ge}$ is the g to e transition frequency, $E_J$ is the Josephson energy, and $E_C$ is the charging energy.

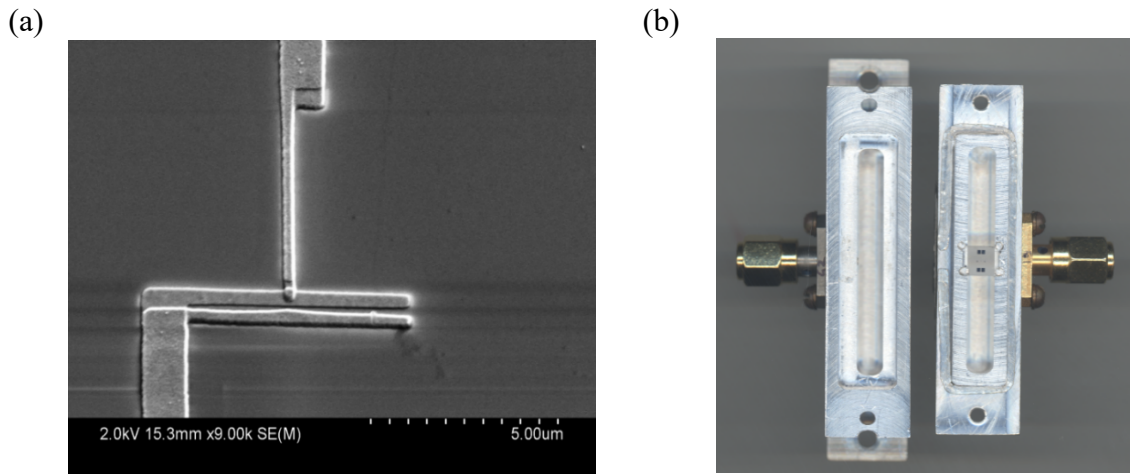

**Fig. 1 (a)** SEM micrograph showing a transmon Josephson junction formed using double-angle evaporation of Al. **(b)** Photograph of the 3D Al cavity, opened up to see the transmon chip mounted in the center of the cavity to cover the left half of the chip, the $O_2$ was evacuated, and we deposited a 65 nm thick layer of nominally pure Al, which formed the counter-electrode of transmon $Q_{R1}$. Patterning concluded with lift-off of the Al layers [see Fig. 1(a)]. The transition temperature and gap of each layer was determined from transport measurements on co-deposited films (see Table 1).

A second chip was fabricated without using the shutter and with both transmons having a $O_2$-doped Al electrode with thickness around 30 nm and a counter-electrode of nominally pure Al around 70 nm thick. For these two devices, labeled $Q_{L2}$ and $Q_{R2}$, the gap values were estimated from our test deposition results and the $O_2$ doping pressure.

Table 1 summarizes the measured parameters for the four transmons. Transmon $Q_{L1}$ had a high-gap top layer and a low-gap bottom layer, while both the top and bottom layers in $Q_{R1}$ had a low gap. For both $Q_{L2}$ and $Q_{R2}$, the top layer had a low gap and the bottom layer had a high gap. In all devices, the top layer had a thickness that was more than twice the thickness of the bottom layer. Note also that all devices except for $Q_{R1}$ had layers with a gap difference that was larger than the transition energy $hf_{ge}$ of the transmon.

After fabrication, each chip was mounted in a 3D aluminum microwave cavity [21] with a fundamental cavity resonance of about 6.115 GHz, as shown in Fig. 1(b). The cavity was mounted to the mixing chamber plate of an Oxford Instruments Triton dilution refrigerator and enclosed in a Cu shield whose inner surface was coated with SiC and epoxy to absorb stray infrared radiation. All microwave lines to the cavity were heavily filtered to block thermally-generated microwave radiation from higher temperatures.

## $T_1$ vs time and $T_1$ vs temperature results

To find the relaxation rate of a transmon, we applied a π-pulse at the device's g-to-e transition frequency $f_{ge}$, waited time $\Delta t$, and then used a high-power pulsed cavity readout technique to measure the state. Repeating this process for a series of $\Delta t$, a complete relaxation curve typically took a few minutes to complete and fitting to this decay curve gave a single value of $T_1$. After the $T_1$ of one transmon was measured, we measured the $T_1$ of the other transmon and repeated this process to get interleaved $T_1$ measurements for both devices on the same chip.

Figure 2(a) shows repeated measurements of the relaxation time $T_1$ of the gap-engineered transmon $Q_{L1}$ and the standard transmon $Q_{R1}$ at the 20 mK base temperature. During the 15 hour span, $T_1$ of the gap-engineered transmon $Q_{L1}$ varied between a minimum of about 100 μs and a maximum of about 310 μs. The corresponding plot of $T_1$ versus time for the standard transmon $Q_{R1}$ showed $T_1$ varying between about 50 and 100 μs. Thus, the gap-engineered device had a $T_1$ that was typically about two or three times longer than that of the standard transmon and both devices showed large fluctuations. Inspection of Fig. 2(a) also reveals that the $T_1$ fluctuations in the two devices were not correlated with each other, even though both transmons were on the same chip and the measurements were interleaved in time. This suggests that the fluctuating loss was local to each qubit.

Figure 2(b) shows repeated measurements of the relaxation time $T_1$ of the gap-engineered transmons $Q_{L2}$ and $Q_{R2}$ at 20 mK. The measurements spanned about 11 hours and large fluctuations in $T_1$ were again obvious in both transmons. The two devices showed a similar behavior. $Q_{R2}$ had a slightly longer average value for $T_1$ and both devices showed fluctuating $T_1$ values between about 100 and 150 μs. The maximum $T_1$ for both devices was over 200 μs.

Although the gap-engineered devices $Q_{L1}$, $Q_{L2}$, and $Q_{R2}$ had somewhat longer $T_1$ values than our standard transmon $Q_{R1}$, they were not greatly longer. Also, compared to $Q_{L1}$, transmons $Q_{L2}$ and $Q_{R2}$ had layers with a larger gap difference, and their low gap layer had a larger volume than their high gap layer. Both of these factors should have led to $Q_{L2}$ and $Q_{R2}$ having longer $T_1$ values than $Q_{L1}$, but this was not the case. One way to check that this loss is dominated by quasiparticle tunneling would be to compare the relaxation time $T_1$ to the quasiparticle charge-parity lifetime, as has been reported by several groups [22-26]. Although the charge dispersion spectra [25,26] of our devices show clear evidence for quasiparticle tunneling, initial attempts to measure the charge-parity lifetime were not successful because our setup did not allow for single-shot measurements.

Figure 3(a) shows measurements of $T_1$ vs temperature $T$ for the gap-engineered transmon $Q_{L1}$. These measurements were taken while the mixing chamber temperature was slowly swept higher and lower over a few cycles. Below 160 mK, $T_1$ was only weakly temperature dependent. This small increase (up to 30%) in $T_1$ as the temperature was reduced below 160 mK may be due to non-equilibrium quasiparticles emptying from the high-gap electrode and accumulating on the low-gap electrode, which would increase $T_1$. Modeling of this behavior [27] shows that this tends to occur at a temperature corresponding to the difference in the gaps of the electrodes. At higher temperatures, $T_1$ decreased rapidly due to loss from thermally-generated quasiparticles. Large fluctuations in $T_1$ are apparent in this data set, with $T_1$ ranging between about 70 μs and 200 μs at low temperatures. Careful examination of this semi-log plot also reveals that the relative size of the fluctuations in $T_1$ do not appear to decrease with temperature. The presence of relative fluctuations in $T_1$ below 160 mK and in the thermal tail that are of about the same size suggests that the same fluctuation mechanism is at work over the full temperature range. Since the relaxation in the thermal tail is clearly dominated by quasiparticles, this behavior is consistent with fluctuations in the quasiparticle loss and not fluctuation of some other loss mechanism.

For comparison, Fig. 3(b) shows the corresponding $T_1$ vs $T$ plot for the standard transmon $Q_{R1}$, which was acquired at the same time as that for transmon $Q_{L1}$. This transmon also showed a

(a)

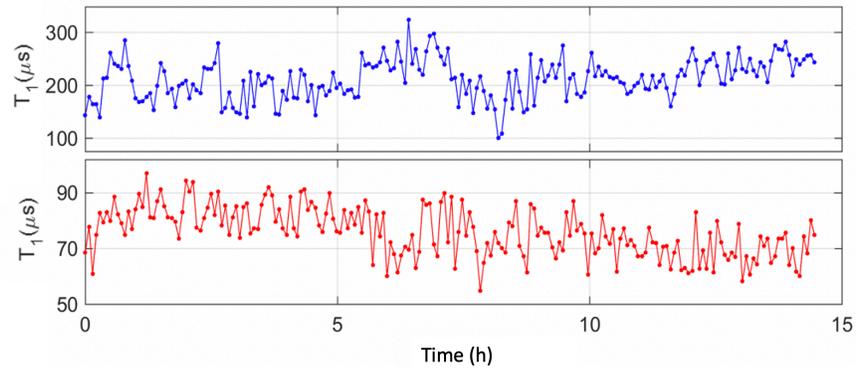

(b)

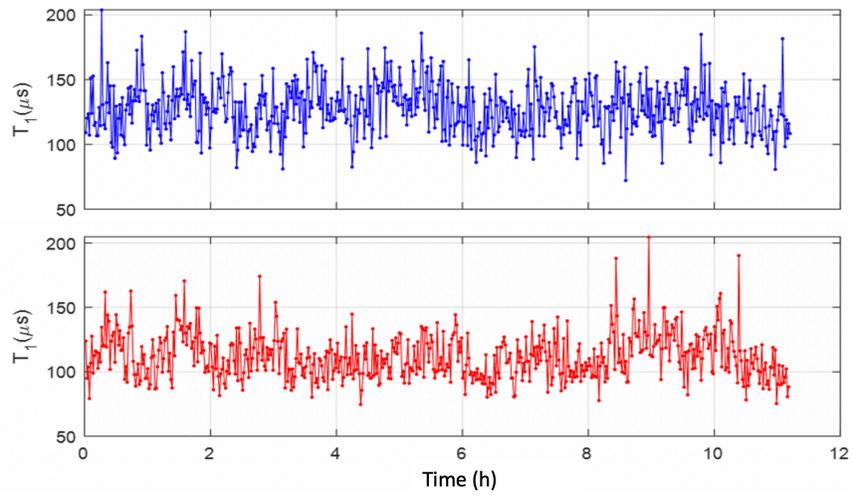

**Fig. 2 (a)** Plot of interleaved $T_1$ vs time data for gap engineered transmon $Q_{L1}$ (blue) and standard transmon $Q_{R1}$ (red), which were on the same chip. **(b)** Plot of interleaved $T_1$ vs time data for transmon $Q_{L2}$ (blue) and transmon $Q_{R2}$ (red)

(a)

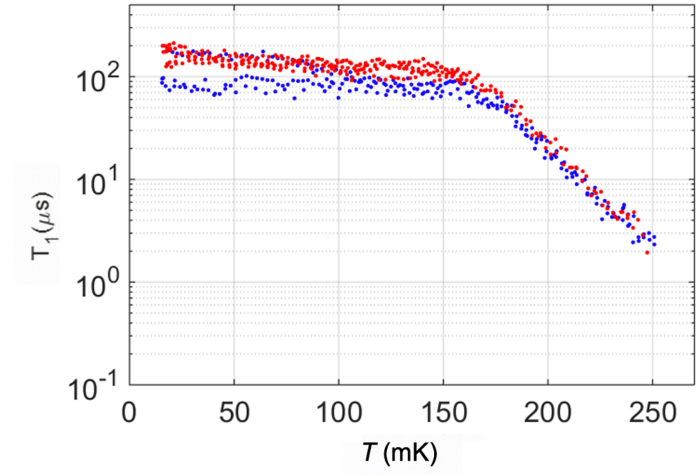

(b)

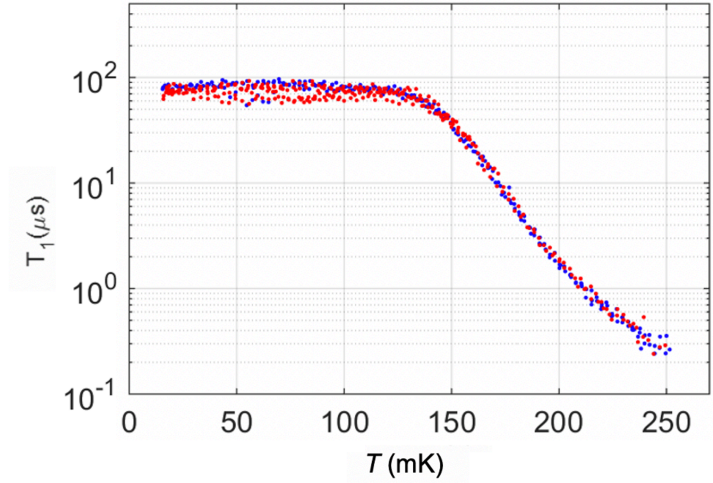

**Fig. 3 (a)** Relaxation time $T_1$ vs temperature $T$ of gap-engineered transmon $Q_{L1}$. Measurements of $T_1$ were acquired while ramping the mixing chamber temperature, with different cycles indicated by red and blue points. Each point took about 5 minutes to acquire. **(b)** $T_1$ vs temperature $T$ of standard transmon $Q_{R1}$

.

relatively temperature-independent $T_1$ below about 140 mK and a rapid decrease in $T_1$ above this due to thermal quasiparticles. Comparing Fig. 3(a) and (b) shows the rapid decrease starts at a somewhat lower temperature in Fig. 3(b), consistent with the smaller gap of the $Q_{R1}$ counter-electrode. Relatively large fluctuations are also evident in this plot, with $T_1$ ranging between about 50 and 100 μs at low temperature.

## Conclusions

In conclusion, we built gap-engineered and standard Al/AlO$_x$/Al transmons. $T_1$ measurements showed large fluctuations in the relaxation time, with $T_1$ ranging as low as about 70 μs up to about 310 μs for gap engineered transmon $Q_{L1}$. $T_1$ varied between about 70 and 200 μs for two other gap-engineered transmons. These $T_1$ values were somewhat longer than that of our standard transmon, but not as long as expected if the loss was dominated by non-equilibrium quasiparticle tunneling across the junction and we were successful at trapping quasiparticles in the low-gap layer. These discrepancies may indicate the presence of other loss mechanisms, such as from two-level systems, or a failure of non-equilibrium quasiparticles to thermalize and empty from the higher-gap electrode. Further experiments, including measurements of the charge-parity lifetime, may help clarify the situation.

## Acknowledgements

We acknowledge the support of the Maryland Quantum Materials Center, the Joint Quantum Institute, and the Laboratory for Physical Sciences and thank the Maryland NanoCenter and FabLab for assistance with device fabrication.

## Conflict of interest

The authors have no conflicts of interest to declare that are relevant to the content of this article.